\documentclass[manuscript,screen]{acmart}

\AtBeginDocument{%
  \providecommand\BibTeX{{%
    \normalfont B\kern-0.5em{\scshape i\kern-0.25em b}\kern-0.8em\TeX}}}

\begin{document}

\title{Position Paper on Simulating Privacy Dynamics in Recommender Systems}

\author{Peter Müllner}
\email{pmuellner@know-center.at}
\affiliation{%
  \institution{Know-Center Gmbh}
  \city{Graz}
  \country{Austria}
}

\author{Elisabeth Lex}
\email{elisabeth.lex@tugraz.at}
\affiliation{%
  \institution{Graz University of Technology}
  \city{Graz}
  \country{Austria}}

\author{Dominik Kowald}
\email{dkowald@know-center.at}
\affiliation{%
  \institution{Know-Center Gmbh}
  \city{Graz}
  \country{Austria}
}

\renewcommand{\shortauthors}{Müllner, et al.}

\settopmatter{printacmref=false}

\begin{abstract}
    In this position paper, we discuss the merits of simulating privacy dynamics in recommender systems.
    We study this issue at hand from two perspectives: Firstly, we present a conceptual approach to integrate privacy into recommender system simulations, whose key elements are \emph{privacy agents}.
    These agents can enhance users' profiles with different privacy preferences, e.g., their inclination to disclose data to the recommender system.
    Plus, they can protect users' privacy by guarding all actions that could be a threat to privacy.
    For example, agents can prohibit a user's privacy-threatening actions or apply privacy-enhancing techniques, e.g., Differential Privacy, to make actions less threatening.
    Secondly, we identify three critical topics for future research in privacy-aware recommender system simulations: (i) \emph{How could we model users' privacy preferences and protect users from performing any privacy-threatening actions?} (ii) \emph{To what extent do privacy agents modify the users' document preferences?} (iii) \emph{How do privacy preferences and privacy protections impact recommendations and privacy of others?}
    Our conceptual privacy-aware simulation approach makes it possible to investigate the impact of privacy preferences and privacy protection on the micro-level, i.e., a single user, but also on the macro-level, i.e., all recommender system users.
    With this work, we hope to present perspectives on how privacy-aware simulations could be realized, such that they enable researchers to study the dynamics of privacy within a recommender system.
\end{abstract}




\maketitle

\section{Introduction}
Through the increasing interest in novel evaluation methods for state-of-the-art recommender systems, research recognized that monitoring a recommender system from a static point of view is often not optimal.
As an alternative, observing the dynamics of a recommender system via simulations showed its merit.
However, modeling the complex and dynamic processes within a recommender system requires lots of engineering efforts.
Hence, several popular simulation frameworks for recommender systems exist~\cite{bountouridis2019siren,ie2019recsim}.
In particular, simulations are beneficial for tracking any kind of effect within a recommender system.
In this regard, recent simulation studies focus on, e.g., the evolution of popularity bias and feedback loops~\cite{yao2021measuring,jiang2019degenerate}, or the confounding of user behavior~\cite{mansoury2020feedback,chaney2018algorithmic}.

In this paper, we present a modified simulation scheme for privacy-aware simulations, which makes it possible to discover the dynamics of privacy within a recommender system.
Specifically, we propose \emph{privacy agents} that enhance existing simulations with the ability to model privacy preferences and protect the privacy of users in a recommender system. 
A privacy agent bridges the gap between a user's preference model~\cite{zhao2019toward} and the user's privacy model~\cite{grace2017towards,mugunthan2020privacyfl}, and this way, the privacy agent serves two purposes.
Firstly, it models a user's privacy preferences, e.g., a user's decision to disclose only certain information to the recommender system.
Secondly, it protects a user from leaking sensitive information by guarding the user's interactions with the recommender system.
For example, when a user selects documents from the recommendation list and gives feedback to the recommender system, sensitive information could get leaked~\cite{ramakrishnan2001being}.
Typically, users have only minor knowledge of privacy threats and are unaware of privacy-enhancing techniques suitable to protect their privacy.
Here, privacy agents provide a remedy, since they intervene and apply state-of-the-art privacy-enhancing techniques, e.g., Encryption~\cite{gentry2009fully} or Differential Privacy~\cite{dwork2014algorithmic}, to protect users from performing any privacy-threatening actions.

Moreover, we state three important issues, that are -- as we think -- crucial for designing privacy-aware recommender system simulations through privacy agents.
Specifically, we briefly discuss (i) how adequate privacy agents could be developed, (ii) to what extent the privacy agents impact users' preferences and thus, drift from reality, and (iii) how one user's privacy agent could positively or negatively influence the recommendations and privacy of other users.

We think that similar to effects like, e.g., popularity bias, also privacy should not only be studied from a static point of view but by monitoring its dynamics in a recommender system.
Thus, with our conceptual privacy-aware recommender system simulation approach presented in this position paper, we aim to give a future perspective on how the important aspect of privacy could be simulated.
We hope that privacy-aware simulations complement insights gained by static evaluations and reveal interesting dynamics of users' privacy within a recommender system.

\section{Current Perspective: Privacy-Agnostic Simulations}
Over the last years, a plethora of research emerged that simulates users and their interactions with a recommender system to grasp the dynamic nature of such systems.
Issues to which recommender system simulations have been applied to include studying effects like item coverage or feedback loops, and analyzing the evolution of user preferences.

In particular,~\cite{ferraro2020exploring,khenissi2020theoretical} show that recommender systems tend to focus on a small set of documents when generating recommendations.
This means that the evolution of the recommender system causes fewer documents to be covered.
Plus,~\cite{khenissi2020theoretical} provides a theoretical framework to model the recommender system's iterative behavior to investigate the presence of different biases.
In the same vein,~\cite{yao2021measuring} analyze the emergence of popularity bias via recommender system simulations.
In particular, the authors discuss the effect of different choice models and the kind of feedback users give to the recommender system on the emergence of popularity bias.
In addition to feedback loops,~\cite{jiang2019degenerate} study filter bubbles and shed light on their effects on the dynamics of user interests.
In summary, recent research utilized recommender system simulations to cover topics like popularity bias, feedback loops, filter bubbles, item discovery, and the evolution of user preferences.
However, we highlight that research works -- to the best of our knowledge -- only use
privacy-agnostic simulations.

\begin{figure}[!htb]
    \centering
    \includegraphics[width=0.9\linewidth]{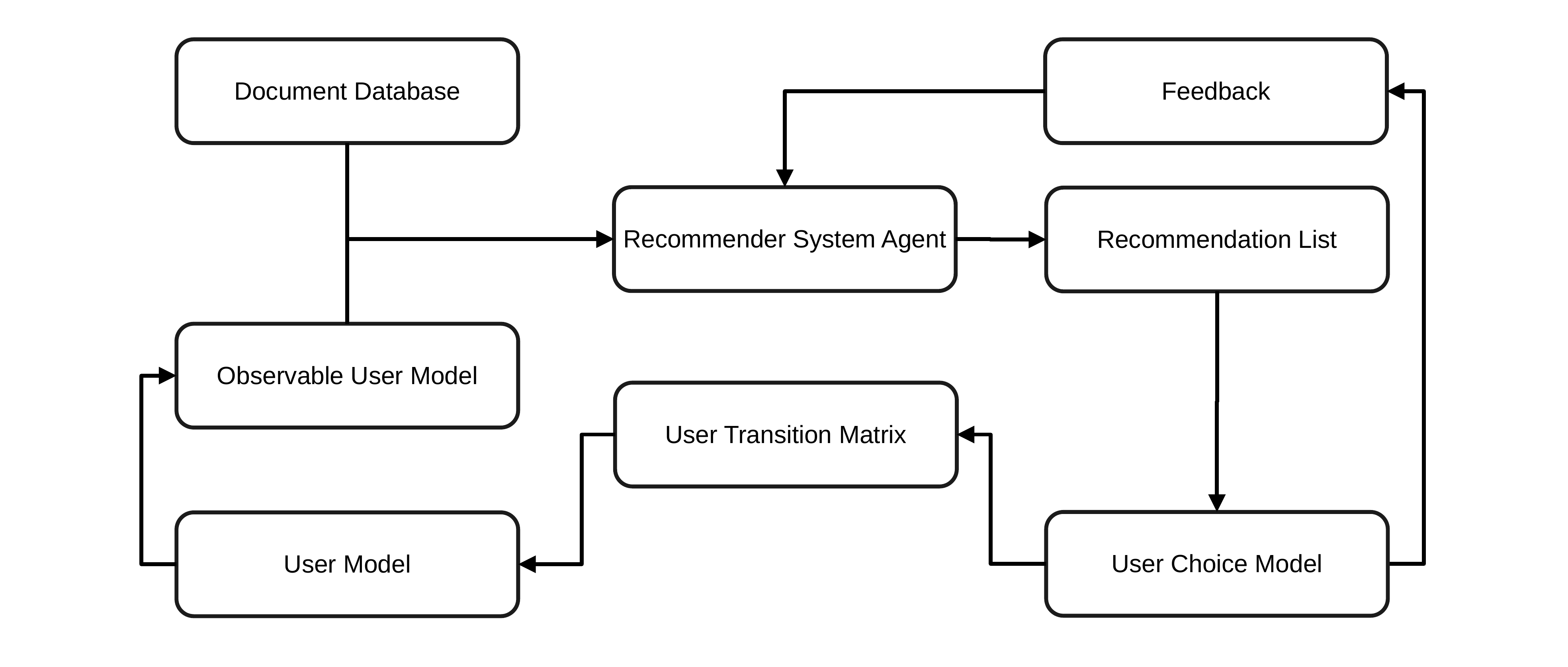}
    \caption{Scheme of a typical privacy-agnostic recommender system simulator (motivated by~\cite{ie2019recsim}). The Recommender System Agent generates a Recommendation List from documents within the Document Database, based on its knowledge (i.e., Observable User Model) about a user (i.e., User Model). Then, the user chooses documents from the Recommendation List via a User Choice Model and gives Feedback to the Recommender System Agent. Plus, the documents chosen from the Recommendation List could lead to transitions in the User Model (i.e., User Transition Matrix).\vspace{-5mm}}
    \label{fig:scheme_privacyagnostic}
\end{figure}

A schematic overview of such a privacy-agnostic recommender system simulator is given in Figure~\ref{fig:scheme_privacyagnostic}, which is motivated by~\cite{ie2019recsim}. 
In the \emph{Document Database}, all documents that are available for recommendation are stored. 
Each user is described by an individual \emph{User Model}, which includes the user's document preferences and rating data.
This is slightly different from the \emph{Observable User Model}, which represents the recommender system's model of the user. 
Thus, this is the recommender system's knowledge about the user, which is a subset of the User Model. 
The \emph{User Choice Model} defines how the user selects documents from the recommended list.
For example, some users could seek serendipitous or novel recommendations, while other users would select documents based on familiarity.
Depending on the received recommendations, users may change their interest in a certain direction and thus, shift the trajectory of the simulation.
Thus, the \emph{User Transition Matrix} measures how a user's preferences change over time.
Finally, the \emph{Recommender System Agent} generates a \emph{Recommendation List} and receives the user's \emph{Feedback}.

\paragraph{Open Issue.} Despite their recent popularity, privacy-agnostic simulations miss the important concept of privacy.
For example, they are unaware of users' concerns and preferences regarding privacy.
Furthermore, the effects of privacy protection, e.g., not disclosing data to the recommender system, are only studied in a static setting~\cite{mullner2021robustness}.
Similar to other factors, we believe that also privacy should be studied from a dynamic perspective.
Hence, privacy-aware simulations are necessary to shed light on the dynamics of users' privacy within a recommender system.

\section{Future Perspective: Privacy-Aware Simulations}
\begin{figure}[!htb]
    \centering
    \includegraphics[width=0.87\linewidth]{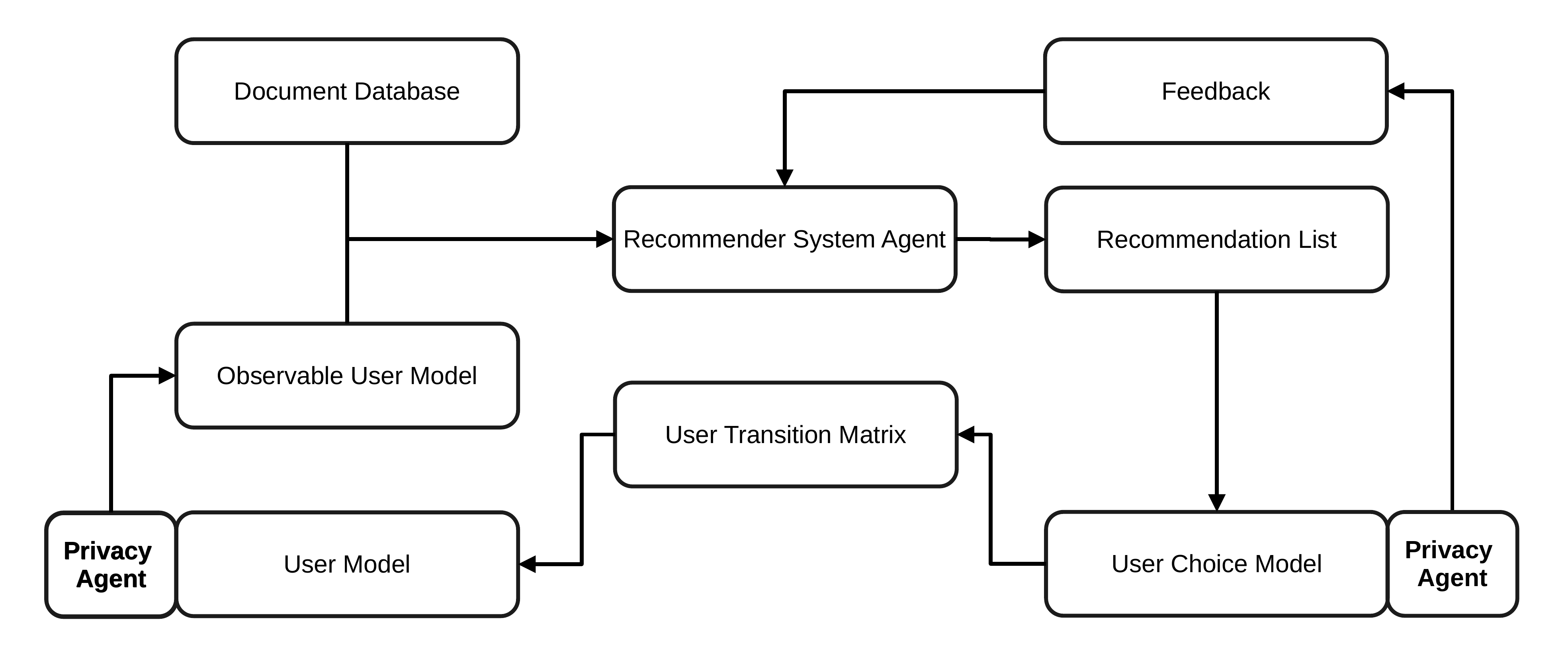}
    \caption{Our proposed scheme of a privacy-aware recommender system simulator. Sensitive data about the user (i.e., User Model) could get leaked through the information available to the recommender system (i.e., Observable User Model) and the user's selection from the Recommendation List as defined by the User Choice Model. We introduce a privacy agent that attaches itself to the User Model and the User Choice Model to (i) model privacy preferences, or (ii) protect privacy.\vspace{-5mm}}
    \label{fig:scheme_privacyaware}
\end{figure}

In this work, we propose privacy agents that enhance the existing recommender system simulation scheme in Figure~\ref{fig:scheme_privacyagnostic} by incorporating users' privacy preferences and privacy protections, and thus, enable privacy-aware simulations.
As illustrated in~Figure~\ref{fig:scheme_privacyaware}, each privacy agent attaches to one user's User Model to (i) model what and how much data the user discloses (i.e., privacy preferences), or (ii) protects the user's privacy by guarding what and how much data is disclosed (i.e., privacy protection).
Furthermore, the privacy agent also attaches to the User Choice Model and (i) models how the user selects documents in line with its privacy concerns (i.e., privacy preferences) or (ii) prohibits the user from selecting documents that would reveal sensitive information (i.e., privacy protection).

\paragraph{Privacy Agent.} 
Firstly, it models a user's privacy preferences, e.g., not disclosing all data to the recommender system.
Secondly, it protects a user's privacy by proposing what data to share and how documents to select from the Recommendation List to not compromise privacy.
For example, the privacy agent could not only model a user's inclination to disclose data to the recommender system but also apply privacy-enhancing techniques to parts of the data, e.g., Encryption~\cite{gentry2009fully} or Differential Privacy~\cite{dwork2014algorithmic}.
In summary, the privacy agent can model users' privacy preferences and, additionally, can protect users from performing actions that would increase the risk of a privacy breach.

\paragraph{Open Research Questions.} With our proposed privacy-aware recommender system simulation scheme, we aim to incorporate privacy to enable more realistic simulations.
However, through the introduction of privacy-aware simulations, several open research questions arise.
Specifically, we highlight three questions that we think are of strong relevance, and thus, worth discussing among researchers.
\begin{enumerate}
    \item \emph{How could we develop privacy agents that model users' privacy preferences, and protect users from performing any privacy-threatening actions?}
    To model users' privacy preferences, one could rely on questionnaires or construct artificial privacy models to model users with different privacy preferences.
    An additional question is how the privacy agents should decide on how to protect users' privacy. 
    In detail, the agents could measure the privacy threat of users' actions by considering the users' privacy preferences but also by measuring the actions' danger from different perspectives (an actions' danger could change depending on the attack scenario), and then apply a suitable privacy-enhancing technique, e.g., Differential Privacy~\cite{dwork2014algorithmic}, Encryption~\cite{gentry2009fully}, or data retention.
    \vspace{3mm}
    \item \emph{To what extent do the privacy agents modify the trajectory of the recommender system? This means, how are users' preferences manipulated when improving privacy?}
    Privacy agents could protect privacy by manipulating the user data available to the recommender system or by manipulating the way the user selects documents from the Recommendation List.
    Since the recommender system relies on these user interactions, the privacy agents influence what recommendations are generated.
    The recommendations can substantially influence users' preferences~\cite{yao2021measuring}, and thus, it follows that protecting privacy by introducing privacy agents could also influence the User Models.
    Therefore, we think it is important to discuss to what extent the privacy agents' impact the User Models and how this could be prohibited.
    \vspace{3mm}
    \item \emph{How does modeling and protecting a user's privacy impact the recommendation quality and privacy of other users in the recommender system?}
    When a privacy agent intervenes to protect a user's privacy, a collaborative recommender system receives less-quality data from this user, which impacts the recommendations of other users within the system.
    This could lead to other users being pushed to disclose more about themselves to receive high-quality recommendations.
    Therefore, (i) monitoring the consequences of users' privacy agents, and (ii) making users aware of the consequences on other users seem to be two interesting aspects worth studying.
\end{enumerate}

\section{Conclusion \& Outlook}
In this position paper at hand, we presented our perspective on privacy-aware recommender system simulations to study privacy in recommender systems from a dynamic point of view.
In particular, we proposed to enhance existing privacy-agnostic simulation schemes by adding a privacy agent for each user.
Firstly, the privacy agents model users' privacy preferences, e.g., their inclination to disclose data to the recommender system.
Secondly, they guard users' interactions with the recommender system and apply privacy-enhancing techniques to protect users from leaking sensitive data through making privacy-threatening actions.
Moreover, we discussed open research questions that are, in our opinion, strongly relevant for establishing privacy-aware simulations via privacy agents.

For future research, we suggest testing our privacy-aware simulation scheme by incorporating privacy agents into different simulation frameworks. 
Furthermore, in analogy to users collaboratively generating recommendations, we propose to research how privacy agents could collaborate to increase recommendation quality and privacy of all users simultaneously (i.e., optimize the privacy-accuracy trade-off).

\paragraph{Acknowledgements.}
This work was supported by the H2020 project TRIPLE (GA: 863420) and the ``DDAI'' COMET Module within the COMET — Competence Centers for Excellent Technologies Programme, funded by the Austrian Federal Ministry (BMK and BMDW), the Austrian Research Promotion Agency (FFG), the province of Styria (SFG) and partners from industry and academia. The COMET Programme is managed by FFG.

\bibliographystyle{ACM-Reference-Format}
\bibliography{sample-base}

\end{document}